\documentclass[journal]{IEEEtran}
\usepackage{cite}
\usepackage{amsmath}
\usepackage{graphicx}
\usepackage{caption}
\usepackage{mathtools}
\usepackage{textcomp}
\usepackage{multicol}
\DeclareUnicodeCharacter{0303}{-}
\graphicspath{{images/}}
\DeclarePairedDelimiter{\ceil}{\lceil}{\rceil}
\newcommand{\Mod}[1]{\ (\mathrm{mod}\ #1)}

\begin{document}
%Header - title and authors
\title{A Novel Reconfigurable Vector-Processed Interleaving Algorithm for a DVB-RCS2 Turbo Encoder}
\author
{Ohad Boxerman\IEEEauthorrefmark{1}, Moshe Bensimon\IEEEauthorrefmark{1}, Shlomo Greenberg, and Yehuda Ben-Shimol
\thanks{\IEEEauthorrefmark{1} M. Bensimon, and O. Boxerman Contribute Equally.
}}
% create the title defined above
\maketitle
%Abstract
\begin{abstract}
Turbo-Codes (TC) are a family of convolutional codes enabling Forward-Error-Correction (FEC) while approaching the theoretical limit of channel capacity predicted by Shannon’s theorem. One of the bottlenecks of a Turbo Encoder (TE) lies in the non-uniform interleaving stage. Interleaving algorithms require stalling the input vector bits before the bit rearrangement causing a delay in the overall process. This paper presents performance enhancement via a parallel algorithm for the interleaving stage of a Turbo Encoder application compliant with the DVB-RCS2 standard. The algorithm efficiently implements the interleaving operation while utilizing attributes of a given DSP. We will discuss and compare a serial model for the TE, with the presented parallel processed algorithm. Results showed a speed-up factor of up to 3.4 Total-Cycles, 4.8 Write and 7.3 Read.
\end{abstract}

%%%%%%%%%%%%%%%%%%%%%%%%%%%%%%%%%%%%%%%%%%%%%%%%%%%%%%%%%%%%
%%%%%%%%%%%%%%%%%%%%%%%%%%%%%%%%%%%%%%%%%%%%%%%%%%%%%%%%%%%%

%\input{Introduction.tex}
\section{Introduction}
\label{sec:intro}
%general introduction to the introduction
Modern age technology and the use of communicative devices dictate the need for efficient communication protocols, data transfers and eminent data manipulation. Nowadays, applications, such as video broadcasting and satellite communications, rely upon fast encoding and calculation tools. Parallelization of data processing requires, in some cases, innovative thinking and different memory access routines unlike the common-default ordered memory structures used. These approaches are the cause of major performance enhancements of many applications. Modern processors use multiple cores for processing data, but in many cases, the program that implements the processing algorithm is sequential and unsuited for the multi-core-vector attributes of the processor, hence not exploiting the full potential of the processor. For these reasons, there is a need for generic tools and techniques for converting known sequential algorithms to a more efficient parallel implementation over vector and multi-core processors existing in today’s market.

% Error correction
\subsection{Error Correction}
Wireless channels impose multiple noise deficiencies that affect the transmitted information. High data transfer rates require fast and efficient encoding that impose both rapid data transmission and eminent data integrity. Data integrity is achieved by data encoding that enables both error detection and correction at the receiver, that is, correction of the received data without the need of resending damaged data packets.

TC implements an approach to the mathematical bounds of data rates with respect to the given noise over given channels, found by C.E. Shannon \cite{6773024,397441}. TC adds redundant bits to the original message to enable Forward Error Correction (FEC), detection and correction of errors at the receiving-end, thus improving data integrity. These techniques are based on the fact that some communication channels suffer from noises over certain segments of the frequency spectrum, and adding redundant bits with a known permutation enables a reliable detection and correction of defected messages with high probability. Additional examples of FEC algorithms include Reed-Solomon codes, Low-Density Parity-Check, Repeat-Accumulate codes and Product codes \cite{abbasfar}.
%TCE
\subsection{Turbo Encoders}
\label{sec:TE}
TC, a subdivision of the Parallel Concatenated Convolutional Codes family are widely used, mainly because of their adjust-ability for the purpose of real-time implementations (e.g. in satellite communications \cite{dvb}). TC are in use in wireless ATMs, Third-Generation systems and in video broadcasting \cite{5763919}. Discussion on the topic could be found in \cite{6325191,BER,901892}. Every TE is constructed by two main operation blocks: an interleaver and a convolution-encoder. While the encoding is the purpose of a TE, interleaving causes substantial operation latency and performance skew. Interleaving is the action of creating a permutation of the input data by a deterministic shuffle. Modulo calculations are a common tool used in programming implementations in order to keep the data in line with a finite vector size or memory allocated array. Modulo calculations are a hardware obstacle which needs to be addressed, especially when dealing with various and large modulo bases. In the TE interleaving block, modulo is a repeated operation consuming both power and computation time. This article presents a vector-oriented implementation of the permutation stage of a specific TE described by the DVBRCS2 standard. Sec.~\ref{sec:ourTE} elaborates the specifications of the turbo code implemented in this research.
\subsection{Vector Processor and Parallel Processing}
\label{sec:VPPP}
A Vector Processor (VP) differs from a Scalar Processor (SP) by its ability to process a single instruction on multiple data (SIMD), usually a one-dimensional array (vector). That 2 is, a single operation is performed simultaneously on N independent elements (N being the given length of the vector). The VP can fetch N elements using a single load operation, thus saving time in both fetching and decoding the data and instructions. In a VP’s architecture it is essential to redefine operations while indicating whether the operation is performed on a scalar or a vector of certain size. The vectors could be of constant/dynamic length. Amongst the applications suitable that can benefit from the accelerated performance on a VP are multimedia, signal processing, cryptography etc. Since the interleaving stage of the TE is a manipulation of long bit-arrays, the choice of a VP is a natural one. Requiring the entire packet of bits for the interleaving stage causes it to be a bottleneck in the TE application. Our simulations revealed that 78\% of processing cycles were of this stage, therefore, improving the performance of the permutation algorithm will result with speedup of the whole application. Using a VP allows to efficiently manipulate the input data by adequate vector and parallel operations, thus enhancing performance of the runtime application both in memory access (R/W) and in total cycles. The algorithmic solution presented in this paper is general and can be reconfigured by the basic DVB-RCS2 parameters.
% Related work
\subsection{Related Work}
\label{sec:relatedWork}
With growing demand for fast data transfers, in recent years we notice an increase in hardware accelerators specifically designed for turbo encoding. A number of hardware accelerators exists: (1) Texas Instruments offers the TCI6618 Multicore SoC DSP with up to 582Mbps throughput. (2) NXP Semiconductors developed the MSC8157 SoC DSP which can reach up to 330 Mbps throughput \cite{nordmark2016turbo}. Algorithms matching vector-abilities and smart SIMD processing with given VP can achieve processing speedup of up to 35\% \cite{kudriavtsev2005generation} while more recent work of comparing vectorized vs. non-vectorized execution achieved up to 57.71\% performance improvement \cite{barik2010efficient}. This article reveals a novel algorithm solving the modulo calculation problem by exploiting the advantages of a VP, thus avoiding the modulo calculation itself. The organization of this paper is as follows: Sec.~\ref{sec:ourTE}, describes the specific TE implemented set by the DVB-RCS2 standard; Sec.~\ref{sec:serial}, describes the original serial permutation algorithm and implementation. This algorithm serves as a reference for the speedup factor calculations; Sec.~\ref{sec:vectoric}, describes the proposed vectorial and parallel permutation algorithm; Sec.~\ref{sec:res}, describes the simulation-based analysis and depicts the results graphically; Finally, Sec.~\ref{sec:con}, concludes the results and achievements and proposes future research possibilities.
\section{The Turbo Encoder}
\label{sec:ourTE}
\begin{figure}[t]
	\centering
	\includegraphics[width = 8.89cm]{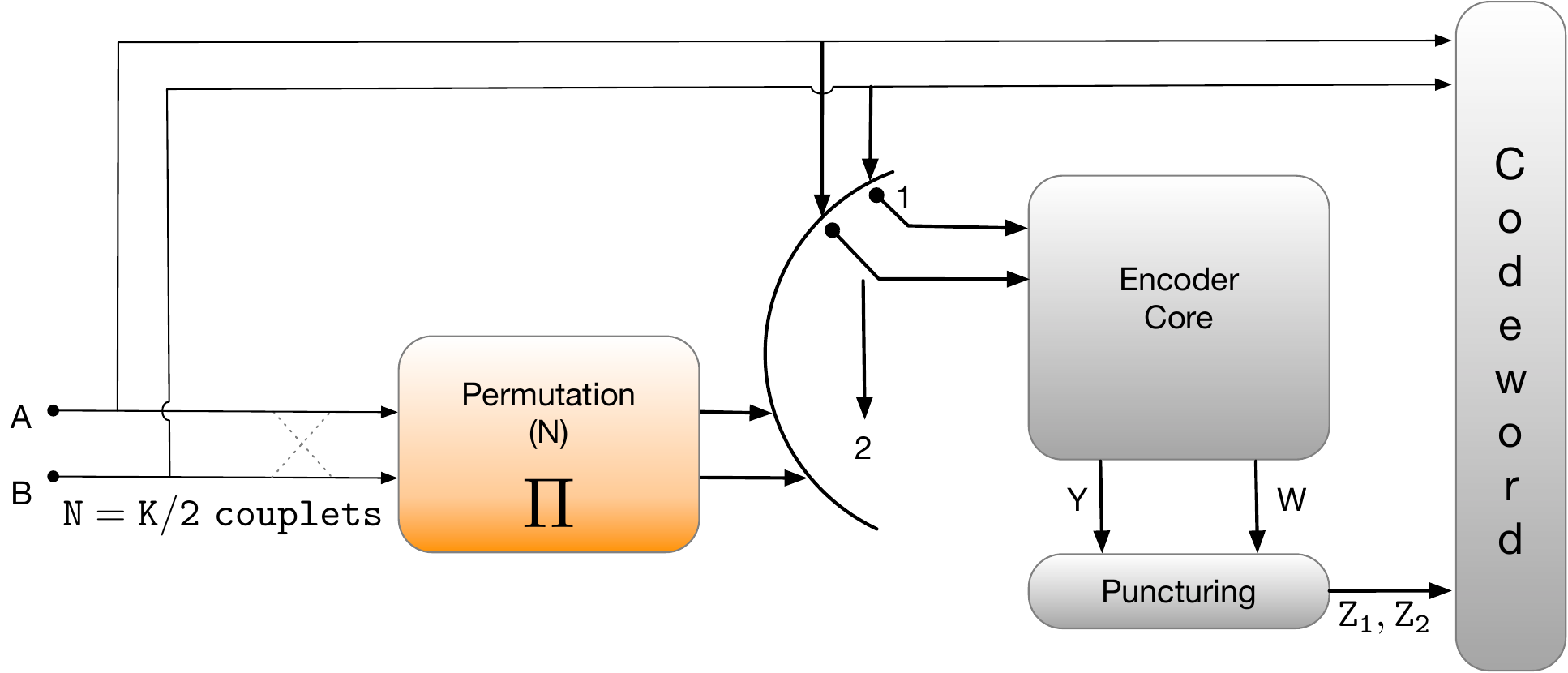}
	\caption{Turbo Encoder Scheme}
	\captionsetup{justification=centering}
	\label{fig:TCEscheme}
\end{figure}

Fig.~\ref{fig:TCEscheme} depicts the TE implemented by the DVB-RCS2 standard. It has a coding rate of 1:3 meaning for every input bit, there are 3 output bits. The input is a 2-bit stream denoted A and B. The output codeword is combined of 3 couplets: (1) The original A and B couple unchanged. (2) A and B are encoded by the ‘Encoder Core’, a convolutional calculation block. (3) N couplets are delayed and rearranged in the ‘Permutation’ block, then encoded via the ‘Encoder Core’ block. 

The permutation stage of the TE is determined by five parameters, $P, Q_0, Q_1, Q_2$ and $Q_3$, with ranges defined by the DVB-RCS2 standard, and vector size N, the number of couplets in bits. The specific parameters used in the implementation are in compliance with the DVB-RCS2 standard and are detailed in Table~\ref{tab:parameters}\cite{dvb}.
\begin{enumerate}
  \item The original A and B couple unchanged. 
  \item A and B are encoded by the `Encoder Core', a convolutional calculation block. 
  \item  $N$ couplets are delayed and rearranged in the `Permutation' block, then encoded via the `Encoder Core' block. 
\end{enumerate}
%Permutation
\label{sec:permute}
The permutation stage of the TE is determined by five parameters, $P, Q_0, Q_1, Q_2$ and $Q_3$, with ranges defined by the DVB-RCS2 standard, and vector size $N$ (the number of couplets in bits). The specific parameters used in the implementation are in compliance with the DVB-RCS2 standard and are detailed in Table~\ref{tab:parameters}\cite{dvb}.

\begin{table}[t]
	\centering
	\caption{Example of used Turbo Encoder parameter-sets}
        % \resizebox{8.3 cm}{!}{	
        \begin{tabular}{c | c | c | c | c | c}
	\hline
		N[bits] & P & $Q_0$ & $Q_1$ & $Q_2$ & $Q_3$\\
		\hline \hline
		56		& 9		& 2		& 2		& 8		& 0\\
		152		& 17	& 9		& 5		& 14	& 1\\
		236		& 23	& 10	& 2		& 11	& 1\\
		384		& 25	& 1		& 2		& 0		& 1\\
		432		& 29	& 1		& 4		& 1		& 1\\
		492		& 31	& 0		& 3		& 1		& 0\\
		520		& 31	& 0		& 1		& 2		& 0\\
		%700		& 37	& 0		& 2		& 0		& 2\\
		776		& 39	& 7		& 0		& 0		& 0\\
		1056	& 43	& 0		& 0		& 6		& 2\\
		1192	& 49	& 0		& 3		& 5		& 0\\
		2396	& 81	& 1		& 2		& 5		& 2\\
		\hline
	\end{tabular}
        % }
	\label{tab:parameters}
\end{table}
Initial simulations running a simple existing serial implementation of this TE resulted in the permutation stage taking up to 78\% of the application (cycle-wise) and showed potential parallelism features such as repetitiveness modulo calculations which could be reconfigured to enhance performance. Fig.~\ref{fig:moduloZoom} depicts an example of permuted indexes by original indexes (for $N = 776$ and matching parameters from Table~\ref{tab:parameters}) and shows the constant $4P$ incrimination leading to the new parallel approach described in Sec.~\ref{sec:vectoric}. The turbo encoder permutation stage is carried out in two levels: 
\begin{figure}[tb]
	\centering
	\includegraphics[width = 8.89cm]{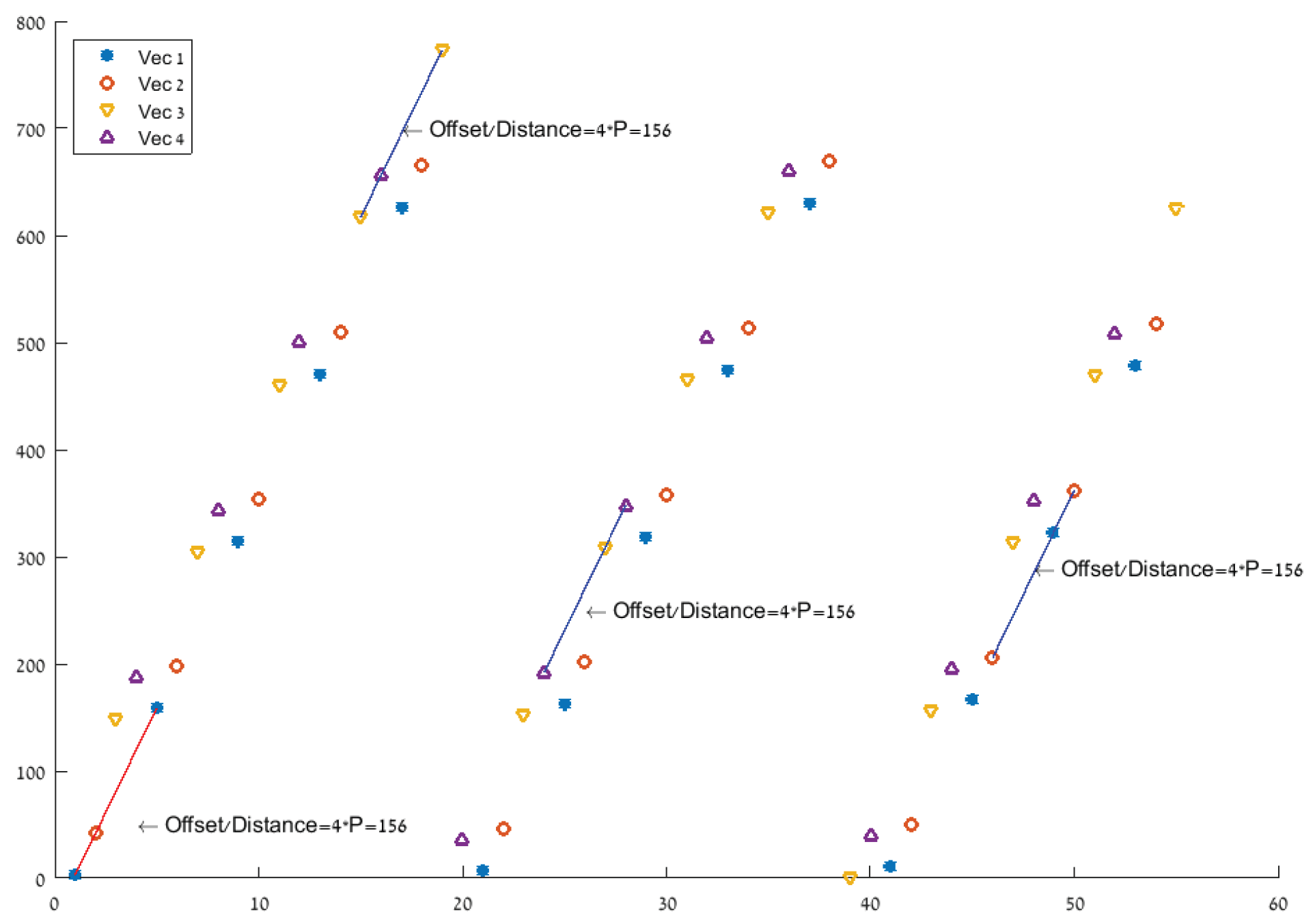}
	\caption{Permuted indexes for example vector length of 776. $\Delta=4\cdot P$ is the offset between two consecutive indexes after permutation.}
	\captionsetup{justification=centering}
	\label{fig:moduloZoom}
\end{figure}
(1) describes a swap of the bit couplets for every odd indexed component

\begin{equation}
\mathrm{if} \{j \Mod{2} = 1 \} \Rightarrow (A_j,B_j) \rightarrow (B_j,A_j),
\label{eq:between}
\end{equation}
(2) describes a calculated rearrangement of the indexes based on the chosen parameters. Further elaboration of the latter can be found in Section IV.
\begin{equation}
i \equiv \Pi(j) = (P\cdot j + Q + 3) \Mod N
\label{eq:within}
\end{equation}
where
\begin{equation}
Q=
\begin{cases*}
0,	&	$j \Mod 4 = 0$ \\
4Q_1,	&	$j \Mod 4 = 1$ \\
4Q_0\cdot P + 4Q_2,	&	$j \Mod 4 = 2$ \\
4Q_0\cdot P + 4Q_3,	&	$j \Mod 4 = 3$ \\
\end{cases*}
\label{eq:cases}
\end{equation}

\section{Serial Implementation}
\label{sec:serial}
The serial implementation serves as reference for speedup and relies on Look Up Tables (LUT). For every input vector, given its length and matching parameters (taken from Table~\ref{tab:parameters}), the program (i) creates LUT by deterministic calculations matching the permutation indexes of each input bit, (ii) loads each bit and its designated new index and (iii) stores the bit in a new output vector. This process is ineffective for two main reasons:
\begin{enumerate}
\item
%%%%%%%%%%%%%%%%%%%%%%%%%%%%%%%%%%%%%%%% add reference to next line%%%%%%%%%%%%%%%%%%%%%%%%%%%%%%%%%%%%
The LUT calculations are determined the DVB-RCS2 standard. The Modulo based calculations are complex for any processor and is a changing parameter disabling simple hardware solutions for these dynamic calculations. Given a finite collection of sets of vector length and parameters, saving all pre-calculated LUT is an option. However, real-time processors run on finite and usually small sized program memory space (MS) bounding the MS that can be reserved for LUT.s
\item
Saving the pre-calculated LUT in the memory doesn’t solve the following issue. Once the LUT are calculated, the program loads bit-by-bit with it’s matching new index from the LUT then the bit is stored to it’s designated index in the output vector. Loading and storing bit-bybit (and index-by-index) results in excessive memory access.
\end{enumerate}
The presented vector algorithm solves these problems by processing the data in max sized streams (loading and storing more than one bit at a time) and by replacing the modulo based calculations of the LUT with data and memory manipulations resulting with lower run-time and memory access. The asymptotic tight bound computational complexity of the serial algorithm is $O(N2)$.
\section{New Approach: Vector Algorithm}
\label{sec:vectoric}
Using a Vector DSP with Very Large Instruction Word (VLIW) architecture requires a different approach compared to the original straightforward approach detailed in Sec.~\ref{sec:serial}. The main idea is based on manipulating the data vectorwise instead of bit-by-bit and taking advantage of the modulo attributes originating from (2) and (3). Fig.~\ref{fig:moduloZoom} indicates modulo induced strides from one index to the next. There are four different groups of indexes derived from the four cases in (3). These strides constitute the basis of the developed vectorwise algorithm described in the following subsections. The algorithm is based on configurable parameters defined by a standard ensuring a comprehensive and generic solution, thus enabling further research possibilities elaborated in Sec.~\ref{sec:con}.

Basing the algorithm on \textit{Load-Execute-Store} operations fitted with the CEVA-XC4500 DSP attributes creates phases. Each phase consists of iterations of execution operations that are done in one work-cycle of the DSP. The following subsections elaborate these operations which combined implement the permutation block of the TE. The presented algorithm was constructed for a specific VP and a specific TE but could be easily modified to various VPs and or different TEs. We chose to test and demonstrate the method on the CEVA-XC4500 and implement the TE described by the DVB-RCS2 standard.
\subsection{Bit to Word}
\label{sec:bit}
\begin{figure}[h]
\centering
\includegraphics[width = 8.89cm]{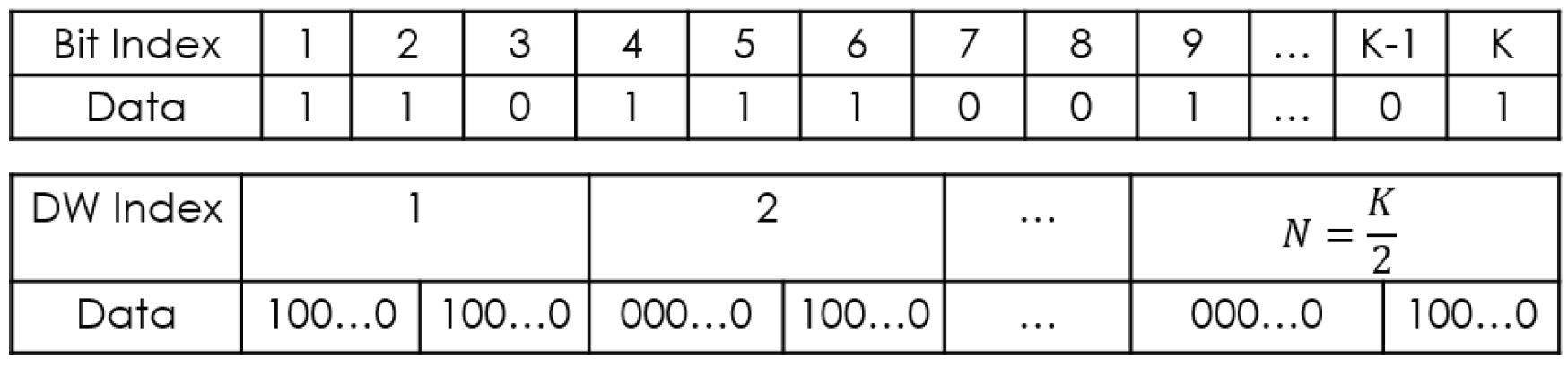}
\caption{Bit to Word - Padding every bit with 15 zeros.}
\captionsetup{justification=centering}
\label{fig:bittoword}
\end{figure}

As in other DSP architectures, the CEVA-XC4500 DSP requires padding of the original input-data for more efficient data manipulation. Hence, the first phase in the algorithm is padding every input-data bit with 15 zeros. Consequently, every original input-data bit is now represented as one Word (16 bits), as shown in Fig.~\ref{fig:bittoword}.

The CEVA-XC4500 DSP is capable of storing 512 bits in one store operation. Following that $16\cdot 32 = 512$, loading 32 bits of the original input-data stream will resolve in 512 bits of padded data as the result of one iteration of this phase. The number of iterations needed for all the original inputdata stream to be processed $\ceil[\big]{\frac{2N}{32}}$ , where N is the number of couplets in bits (see Fig.~\ref{fig:TCEscheme}). Hence, the output vector of this phase is of size $16\times 2N [bits]$ or $2N [words]$. Notes:

The CEVA-XC4500 DSP is capable of storing 512 bits in one \textit{store} operation. Following that $16\cdot 32 = 512$, loading 32 bits of the original input-data stream results with 512 bits of padded data in one iteration of this phase. Therefore, the number of iterations needed for all the original input-data stream to be processed is $\ceil[\big]{\frac{2N}{32}}$, where $N$ is the number of couplets in bits (see Fig.\ref{fig:TCEscheme}). Hence, the output vector of this phase is of size $16\times 2N [bits]$ or $2N [words]$.

Notes:
\begin{enumerate}
\item
The permutation stage is identical to every two bits of original input couplet (A and B) and so we refer to every padded couplet as a Double-Word (DW). In addition, The architecture of the CEVA-4500 DSP is designed to processes DWs, therefore the next two phases of permutation are performed on $N$ DWs.
\item
The Bit-to-Word phase is added as an input rearrangement requirement of the specific DSP implementation. 4 Other processors will/will not require different input rearrangements which will affect results.
\end{enumerate}
\subsection{Transpose}
\label{sec:trans}
\begin{figure}[t]
\centering
\includegraphics[width = 8.89cm]{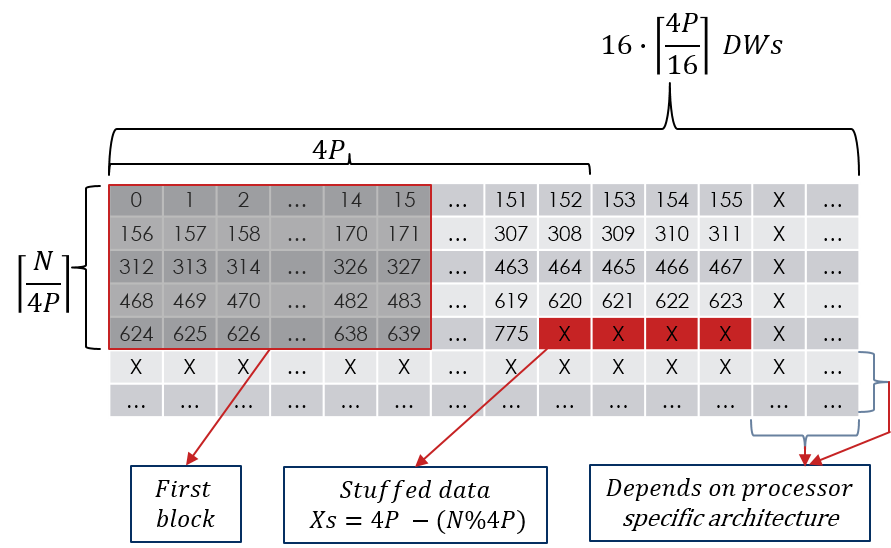}
\caption{Example of a virtual 2D matrix for: $N = 776; P = 39$.}
\captionsetup{justification=centering}
\label{fig:2Dmat}
\end{figure}

\begin{figure}[t]
\centering
\includegraphics[width = 8.89cm]{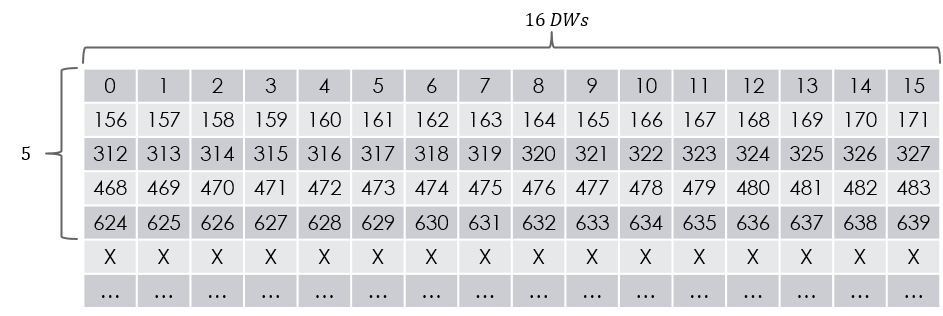}
\caption{The first block of the virtual 2D matrix (highlighted in
Fig.~\ref{fig:2Dmat}) before transposition.}
\captionsetup{justification=centering}
\label{fig:block}
\end{figure}

\begin{figure}[t]
\centering
\includegraphics[width = 6.5cm]{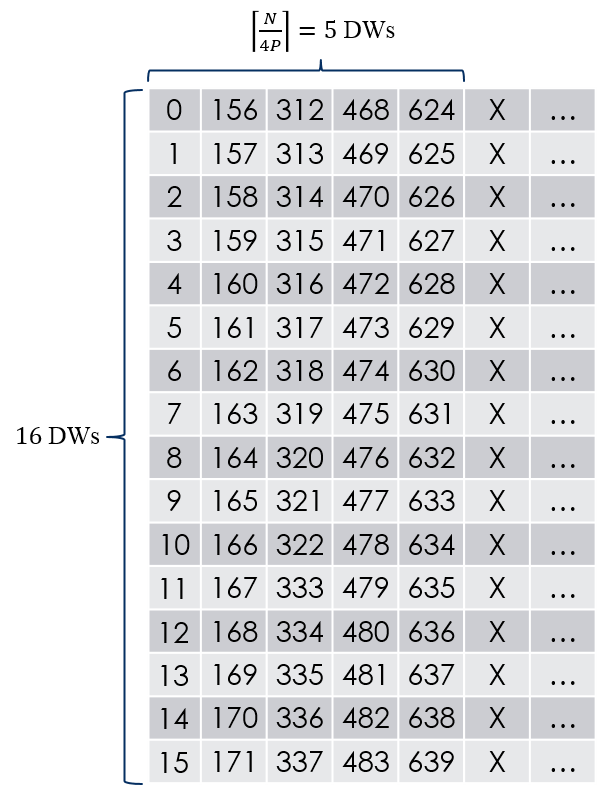}
\caption{Block in Fig. \ref{fig:block} after transposition.}
\captionsetup{justification=centering}
\label{fig:transblock}
\end{figure}

Referring to the input DW vector as a virtual 2D matrix of $\left[\ceil[\big]{\frac{N}{4P}}\right]\times [4P]$, the vector is accessed with constant strides defined by the number of columns $[4P]$ as seen in Fig.~\ref{fig:2Dmat}. This virtual 2D matrix is transposed such that indexes $j +4P$ are the successors of indexes $j$. The CEVA-XC4500 DSP best transposes blocks of $4\times 16 [DWs]$, therefore a block of 4 rows, each of $16 [DWs]$, is loaded from the memory. Fig.~\ref{fig:block} shows an example first block of the virtual 2D matrix illustrated in Fig.~\ref{fig:2Dmat}. The $i^{th}$ block to be transposed is defined by four rows of 16 consecutive DWs. The $k^{th}$ row of block $i$ starts at:
\begin{equation*}
	16i + k\cdot 4P; k = 0,1,2,3, 1\leq i \leq \ceil[\bigg]{\frac{4P}{16}}
	\label{eq:row}
\end{equation*}
and $\ceil[\big]{\frac{4P}{16}}$ is the number of blocks to be transposed. The number of stuffed rows in the last block is $16\cdot \ceil[\big]{\frac{4P}{16}} - 4P$.
Fig.~\ref{fig:transblock} shows the same block illustrated in Fig.~\ref{fig:block}. Each row is stored in a continuous manner. The next row is stored at the end of the previous row. The next block rows are stored at the end of the previous block.
Notes:
\begin{itemize}
	\item 
         An important part of this phase is swapping the words of odd-indexed ($1,3,5,...,N-1$) DWs, meaning that in every odd-indexed DW the high-word and the low-word are swapped. This swapping implements the permutation inside the bit-couplet, as required. The CEVA-XC4500 DSP can swap word while loading any DW, meaning the bit-level swapping is implemented without influencing processing time or memory access.
	\item
        The chosen vector size and parameter sets used in simulations require transposing matrices of up-to eight rows. An expansion of this algorithm for larger sized matrices might be possible but was not simulated.
\end{itemize}

\subsection{Concatenation}
\label{sec:concat}
This phase creates four vectors by concatenating rows of the transposed 2D matrix of the previous phase. The four initial rows are chosen by four pre-calculated offsets determined by the DVB-RCS2 standard parameters $Q_{0}, Q_{1}, Q_{2}, Q_{3}$ and $P$:
% \begin{equation}
% \mathrm{offset}_{i}=
% 	\begin{cases}
% 	3  &  i=1\\
% 	(3+4Q_{1} + P)\Mod N & i=2\\
% 	(3+4(Q_{0}\cdot P + Q_{2})\Mod N + 2P) & i= 3\\
% 	(3+4(Q_{0}\cdot P + Q_{3})\Mod N + 3P) & i=4
% 	\end{cases}
% \end{equation}
\begin{itemize}
	\item $\mathrm{offset}_{1} = 3$
	\item $\mathrm{offset}_{2} = (3+4Q_{1} + P)\Mod N$
	\item $\mathrm{offset}_{3} = (3+4(Q_{0}\cdot P + Q_{2})\Mod N + 2P)$
	\item $\mathrm{offset}_{4} = (3+4(Q_{0}\cdot P + Q_{3})\Mod N + 3P)$
\end{itemize}
The strides taken from one row to the next are determined by $G = mod_{N}(4P \cdot\ceil[\big]{\frac{N}{4P}})$. We continue and construct each vector concatenating the row G from it. Knowing there are $4P$ rows in the transposed 2D matrix and 4 vectors, consequently there are $P-1$ concatenations to be done:
\begin{figure}[h]
	\centering
	\includegraphics[width = 8.89cm]{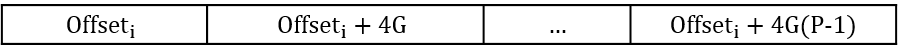}
	\caption{Concatenated $\mathrm{Vector}_i; 1\leq i\leq 4$.}
	\captionsetup{justification=centering}
	\label{fig:concat_vectori}
\end{figure}
\textit{Note}: In Fig. \ref{fig:concat_vectori} all row numbers must wrap around, keeping the index in range: $0\leq row_{idx} \leq 4\cdot (P-1)$.
Some parameter sets dictate a starting point for concatenation that is not the first DW of the first chosen row. The first DW for each $\mathrm{vector}_i$ is defined by:
\begin{equation*}
L_{k} = \ceil[\bigg]{\frac{\mathrm{offset}_{k}}{4P}}; k = 1,2,3,4
\end{equation*}
\begin{figure}[h]
\centering
\includegraphics[width = 8.89cm]{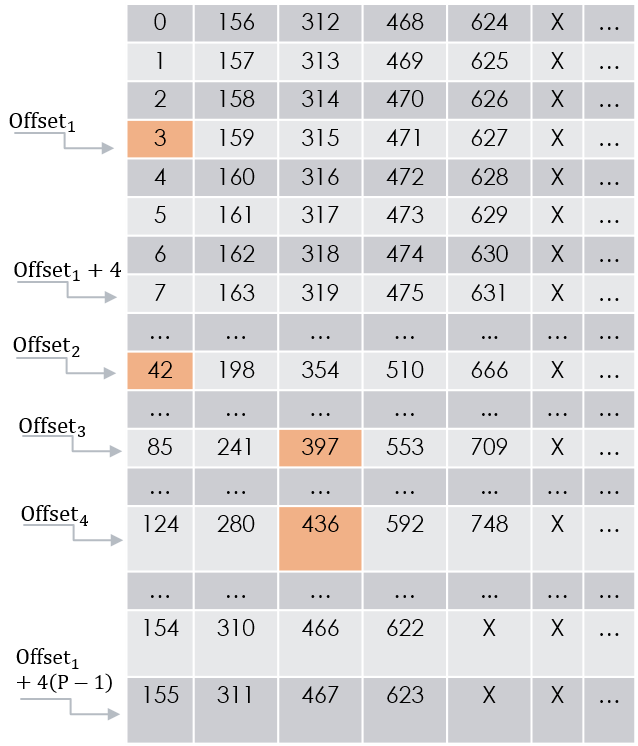}
\caption{Transposed 2D matrix (in Fig. \ref{fig:2Dmat}) with offsets}
\captionsetup{justification=centering}
\label{fig:offset_mat}
\end{figure}
Fig. ~\ref{fig:offset_mat} depicts the virtual 2D matrix illustrated in Fig. \ref{fig:2Dmat}. In this example $N = 776; P = 39; Q_{0} = 7; Q_{1} = 0; Q_{2} = 0; Q_{3} = 0$. Hence, using the expressions above one may obtain: $\mathrm{offset}_{1} = 3$; $\mathrm{offset}_{2} = 42$; $\mathrm{offset}_{3} = 397$; $\mathrm{offset}_{4} = 436$; $G = 4$; $L_{1} = 0$ (always true since $\mathrm{offset}_{1} \equiv 3$); $L_{2} = 0$, $L_{3} = 2$; $L_{4} = 2$.
\begin{figure}[h]
\centering
\includegraphics[width = 8.89cm]{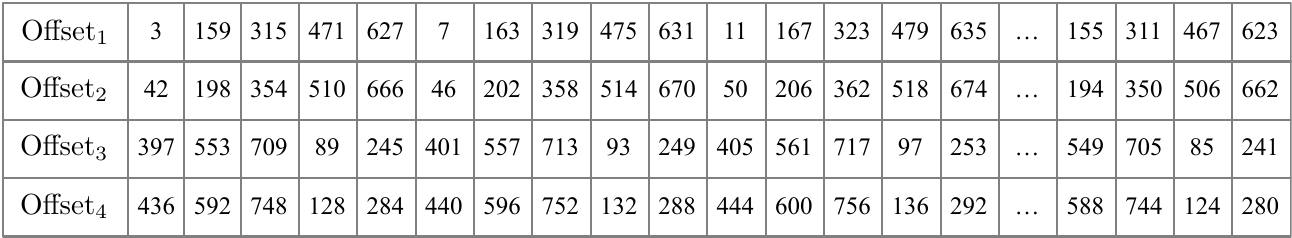}%convecs_noback.png
\caption{Four concatenated vectors (generated from Fig. \ref{fig:offset_mat}).}
\captionsetup{justification=centering}
\label{fig:convecs}
\end{figure}
Fig. ~\ref{fig:convecs} shows an example of four result vectors in positions $\mathrm{offset}_1-\mathrm{offset}_4$ which are the output of the concatenation phase and serve as the input for the ordering phase.
\subsection{Ordering}
\label{sec:order}
The ordering phase organises it's input as one long vector. This is done by loading the four vectors and transposing them using a similar transposition process to the one described in the transpose phase. Here, every iteration transposes blocks of $4\times 16$ [DWs] with the following differences:
\begin{enumerate}
	\item The rows transposed are now stored in an orderly fashion which match the desired output vector (no need for reconstruction of the transposed matrix).
	\item The vectors are of same size $(\frac{N}{4})$, meaning that stuffed data will exist only to fill the last block transposed.
\end{enumerate}
\begin{figure}[h]
\centering
\includegraphics[width = 8.89cm]{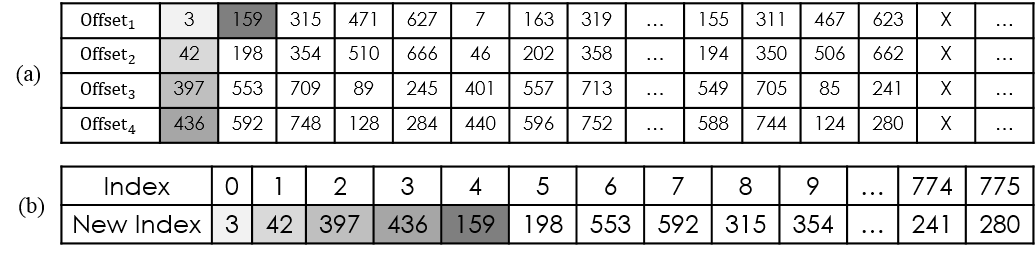}
\caption{Four concatenated vectors and matching output vector for $N = 776; P = 39$.}
\captionsetup{justification=centering}
\label{fig:ordering}
\end{figure}
Therefore, the number of blocks transposed is $\ceil[\big]{\frac{N}{4\cdot 16}}$, and the number of stuffed columns in the last block is $16\cdot\ceil[\bigg]{\frac{N}{4\cdot 16}} - \frac{N}{4}$. Fig. \ref{fig:ordering} illustrates the $4\times \frac{N}{4}$ [DWs] matrix before reordering and the output of this phase, $N$ [DWs] after permutation (example).
\subsection{Word To Bit}
\label{sec:word}
The final operation of the algorithm is un-padding the output vector of the ordering phase from the redundant zeros that were added to the original data in the ”bit to word” phase. 

This concludes the vector-wise permutation stage implementation. Given $N$ bit couplets and matching parameters, following these phases, the result vector will contain $N$ bit couplets after permutation. The described are compatible with the DVB-RCS2 standard and are as generic as possible, having little data adjustments to best-fit CEVA’s DSP can be reconstructed to fit any VP with very few and minor changes to the overall algorithm. 

The asymptotic tight bound computational complexity of the vector-wise algorithm can be written as: 
\begin{equation*}
	O(N+\ceil[\big]{\frac{N}{P}}\cdot P) \simeq O(N)
\end{equation*}
\section{Simulations And Results}
\label{sec:res}
Performance analysis was carried out using the CEVATOOLBOX profiler designed for CEVA’s DSPs. The profiler creates a table detailing (1) the cycle count of each function in the code and total cycle count results and (2) the number of \textit{Read}/\textit{Write} operations performed in each function and total \textit{Read}/\textit{Write} results. Unlike the approach of enhancing each function of a code/algorithm individually, the proposed vector implementation revised the whole permutation process. Hence, serial and vector algorithms are fundamentally incompatible, therefore we consider only the overall results of the permutation stage detailed in Tables~\ref{tab:profilerRW} and sec.\ref{tab:profilerTC}.

The simulations were executed for 11 vector sizes and matching parameter sets given in Table I. for each parameter set, we compared the serial and vector-wise results and calculated the speedup using:
\begin{equation*}
Speedup = \frac{L_{old}-L_{new}}{L_{new}}\cdot 100\%.
\end{equation*}
where $L_{old}$ and $L_{new}$ are the counts of Read, Write and Total Cycles of the serial and vector-wise implementations, respectively. 

% read/write results
\begin{table}[t]
	\centering
	\caption{Read/Write Profiler Results}
	\resizebox{8.3 cm}{!}{
	\begin{tabular}{ c | c c | c c }
		\hline
		Vector Size & \multicolumn{2}{c |}{Parallel}	& \multicolumn{2}{c}{Speedup}\\
		N[bits] & Read 	& Write	& Read 	& Write	\\
		\hline \hline
		56		& 761	& 666 & 11.7 & 109.6 \\
		152		& 621	& 570 & 82.1 & 158.6 \\
		236		& 739	& 802 & 200.0 & 143.8 \\
		384		& 822	& 719 & 284.7 & 230.4 \\
		432		& 988	& 1042 & 383.8 & 199.0 \\
		492		& 1279	& 997 & 623.4 & 409.5 \\
		520		& 998	& 1042 & 438.5 & 222.2 \\
		776		& 1217	& 928 & 589.9 & 406.4 \\
		1056	& 1487	& 1109 & 307.1 & 229.8 \\
		1192	& 1487	& 1109 & 734.5 & 484.4 \\
		2396	& 1612	& 1189 & 295.2 & 219.6 \\
		\hline
	\end{tabular}
	}
	\label{tab:profilerRW}
\end{table}
% total cycles results
\begin{table}[h]
	\centering
	\caption{Total Cycles Profiler Results}
	\resizebox{8.3 cm}{!}{
		\begin{tabular}{ c | c | c}
			\hline
			Vector Size & Parallel	& Speedup\\
			N[bits] & Total Cycles & Total Cycles\\
			\hline \hline
			56		& 172239 & 6.000\\
			152		& 132264 & 63.70\\
			236		& 175387 & 87.70\\
			384		& 178026 & 141.7\\
			432		& 238599 & 153.1\\
			492		& 270537 & 296.2\\
			520		& 238599 & 177.8\\
			776		& 255186 & 284.1\\
			1056	& 313108 & 134.9\\
			1192	& 313108 & 339.9\\
			2396	& 339651 & 124.7\\
			\hline
		\end{tabular}
	}
	\label{tab:profilerTC}
\end{table}

\begin{figure}[tb]
\centering
\includegraphics[width = 8.89cm]{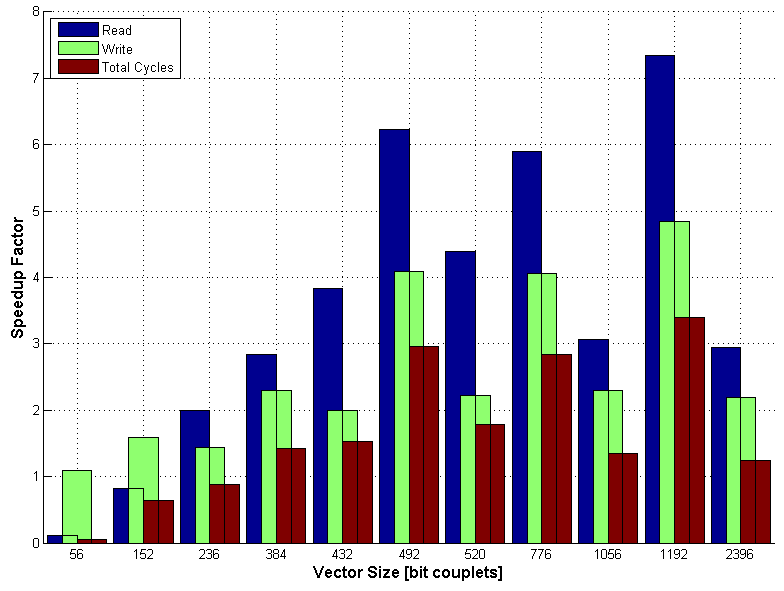}
\caption{Speedup results of Read, Write and Total Cycles by Vector Size}
\captionsetup{justification=centering}
\label{fig:results}
\end{figure}

The results of Tables~\ref{tab:profilerRW} and \ref{tab:profilerTC} are also presented in Fig. ~\ref{fig:results}.
We notice a positive speed-up factor for all vector sizes. The reason for the non-monotonic plots is the different execution of the vector algorithm caused by the given parameters of the each vector size and not only by the vector size itself. Re-analyzing the vector algorithm phases by the parameters we see that for example the number of blocks transposed is a factor of P and can shorten/lengthen the runtime of the application by reducing/adding blocks to transpose. See Sec.~\ref{sec:con} for more.
\section{Conclusion and Future Research}
\label{sec:con}
\subsection{Conclusion}
The vector-wise algorithm implementation out performs the serial implementation for all tested cases. Speedups were achieved even though the DSP hardware required padding and unpadding of the original data and final result respectively. A hardware better fit for the parallel algorithm, enabling the \textit{Transpose, Concatenation and Ordering} phases to be executed on bits rather than words will improve the superiority of the parallel algorithm over the serial implementation.
\subsection{Future Research}
Research possibilities include different paths:
\begin{enumerate}
	\item 
	Additional permutation algorithms can and be tested on the DSP.
	\item
	The parallel algorithm can be implemented or simulated on other processors (e.g. GPUs, CPUs, etc.).
	\item
        The encoding performed by this application is targeted at achieving maximum channel capacity with minimum bit error rate (BER). The parameters used in this research (shown in Table \ref{tab:parameters}) were given as constants. A wider approach would be to optimize vector length with matching parameters, while considering the algorithm presented and it’s vector properties, all this while striving to achieve better BER and throughput.	
\end{enumerate}

%\section{acknowledgment}
%\label{sec:ack}
%We would like to thank the HiPer Consortium and the CEVA and SatixFy companies for their support in this research.

%\section{References}
\bibliographystyle{IEEEtran}
\bibliography{Permutation_Paper}
\end{document}